\documentclass[12pt]{article}
\usepackage{times}
\usepackage{color}
\usepackage{float}

\usepackage{hyperref}
\usepackage{academicons}
\usepackage{xcolor}
\newcommand{\orcid}[1]{\href{https://orcid.org/#1}{\textcolor[HTML]{A6CE39}{\aiOrcid}}}

\DeclareRobustCommand{\ion}[2]{%
\relax\ifmmode
\ifx\testbx\f@series
{\mathbf{#1\,\mathsc{#2}}}\else
{\mathrm{#1\,\mathsc{#2}}}\fi
\else\textup{#1\,{\mdseries\textsc{#2}}}%
\fi}

\newcommand{\apj}{Astrophys. J.}
\newcommand{\apjl}{Astrophys. J. Lett.}
\newcommand{\apjs}{Astrophys. J. Supp.}
\newcommand{\mnras}{Mon. Not. R. Astron Soc.}
\newcommand{\aap}{Astron. Astrophys.}

\newcommand{\solphys}{Sol. Phys.}
\newcommand{\ssr}{Space Sci. Rev.}
\newcommand{\pop}{Phys. Plasmas}
\newcommand{\prl}{Phys. Rev. Lett.}
\newcommand{\nat}{Nature}

\newcommand{\araa}{Annual Review of Astronomy and Astrophysics}

\title{The Origin of Underdense Plasma Downflows Associated with Magnetic Reconnection in Solar Flares}

\author
{Chengcai Shen$^{1\ast}$, Bin Chen$^{2}$, Katharine K. Reeves$^{1}$, Sijie Yu$^{2}$, Vanessa Polito$^{3,4}$, \\
Xiaoyan Xie$^{1,5,6}$
\\
\\
\normalsize{$^{1}$Center for Astrophysics \textbar \ Harvard \& Smithsonian,}\\
\normalsize{60 Garden Street, Cambridge, MA, 02138, USA}\\
\normalsize{$^{2}$Center for Solar Terrestrial Research, New Jersey Institute of Technology,}\\
\normalsize{University Heights, Newark, NJ 07102, USA}\\
\normalsize{$^{3}$Bay Area Environmental Research Institute, NASA Research Park, Moffett Field, CA, USA}\\
\normalsize{$^{4}$Lockheed Martin Solar and Astrophysics Laboratory, Building 252,}\\
\normalsize{3251 Hanover Street, Palo Alto, CA 94304, USA}\\
\normalsize{$^{5}$Yunnan Observatories, Chinese Academy of Sciences, Kunming, Yunnan 650216, China}\\
\normalsize{$^{6}$University of Chinese Academy of Sciences, 19A Yuquan Road, Beijing 100049, China} \\
\\
\normalsize{$^\ast$Correspondence E-mail: chengcaishen@cfa.harvard.edu.}
}

\date{}

\usepackage{graphicx}

\begin{document} 


\baselineskip24pt

\maketitle

\section*{}
{\bf 
Magnetic reconnection is a universal process that powers explosive energy release events such as solar flares, geomagnetic substorms, and some astrophysical jets. A characteristic feature of magnetic reconnection is the production of fast reconnection outflow jets near the plasma Alfv\'{e}n speeds \cite{Forbes1996ApJ...459..330F,2018PhPl...25j2120H}.
In eruptive solar flares, dark, finger-shaped plasma downflows moving toward the flare arcade have been commonly regarded as the principal observational evidence for such reconnection-driven outflows\cite{Mckenzie1999ApJ...519L..93M, Savage2012ApJ...747L..40S}. However, they often show a speed much slower than that expected in reconnection theories \cite{savage_quantitative_2011,innes_observations_2014}
, challenging the reconnection-driven energy release scenario in standard flare models. Here, we present a three-dimensional magnetohydrodynamics model of solar flares.
By comparing the model-predictions with the observed plasma downflow features, we conclude that these dark downflows are
self-organized structures formed in a turbulent interface region below the flare termination shock where the outflows meet the flare arcade, a phenomenon analogous to the formation of similar structures in supernova remnants. This interface region hosts a myriad of turbulent flows, electron currents, and shocks, crucial for flare energy release and particle acceleration.
}

In eruptive solar flares, various plasma flows above the post-reconnection flare arcades have been frequently reported in the literature\cite{Innes2003SoPh..217..267I,Lin2005ApJ...622.1251L,Liu2013ApJ...767..168L,warren_spectroscopic_2018}.
In particular, for flare current sheets that have a face-on viewing geometry (see Figs.~1(A) and (C) for examples), rapidly descending, dark finger-like features, usually referred to as supra-arcade downflows (SADs), have been often observed in extreme ultraviolet (EUV) and soft X-ray (SXR) images. These SADs are embedded in a diffuse fan-like structure above the post-reconnection flare arcades.
Owing to their close resemblance to the morphology and dynamics of the predicted reconnection outflows in magnetohydrodynamics (MHD) simulations, SADs are commonly interpreted as the signature of such outflows residing in the large-scale current sheet, although the detailed mechanism responsible for their dark appearance is under continued debate\cite{Savage2012ApJ...747L..40S,Cassak2013ApJ...775L..14C}.
Their much slower speeds ($\leq $15\% of the Alfv\'{e}nic speed\cite{savage_quantitative_2011}) are discussed in numerical models containing the bursty jet\cite{cecere_3d_2015} or Rayleigh–Taylor-type instabilities\cite{Guo2014ApJ...796L..29G} in reconnection downflow regions. However, several observational thermal features of SADs challenge the above model predictions\cite{Reeves2017ApJ...836...55R}. 
Another possible interpretation involves reconnection outflows that are slowed down by the aerodynamic drag force along their path (which arises when a fast-moving structure cuts through a quasi-static medium)\cite{Longcope2018},
yet observations of the SADs often show that they  flow downward at a nearly constant speed with no sign of appreciable deceleration until they reach the top of the flare arcade \cite{savage_quantitative_2011}.

One important feature that is often overlooked is the ``interface'' region where the downwards fast reconnection outflows impinge upon the closed flare arcades. 
This interface region has also been often referred to as the ``cusp'' (from the appearance of the highly bent field lines viewed edge on; Fig.~1B) or the ``above-the-looptop'' region\cite{1994Natur.371..495M}.
Recently, it has been suggested that this interface region may play a dominant role in flare energy release, particle acceleration, and plasma heating\cite{2010ApJ...714.1108K,2020NatAs...4.1140C,2020Sci...367..278F}.
The violent impact between reconnection downflows and closed flare arcades may induce a variety of physical processes including fast- and slow-mode shocks \cite{Forbes1996ApJ...459..330F,Chen2015Sci...350.1238C,takasao_magnetohydrodynamic_2015}, collapsing magnetic traps\cite{1997ApJ...485..859S}, waves\cite{Reeves2020ApJ...905..165R}, or turbulence \cite{2020Sci...367..278F}.
This interface region above the flare arcade is analogous to the highly turbulent region sandwiched between the forward and reverse shock in supernova remnants, which hosts a variety of instabilities that enable the plasma to develop distinctive, finger-like structures\cite{Miles2009}. A well-known example of such finger-like structures are those observed in supernova remnants\cite{Warren2005}, which share a similar appearance to SADs observed in solar flares (see Fig. 1C, D for a comparison). 

\begin{figure}[H]
    \centering
    \includegraphics[width=1.0\textwidth]{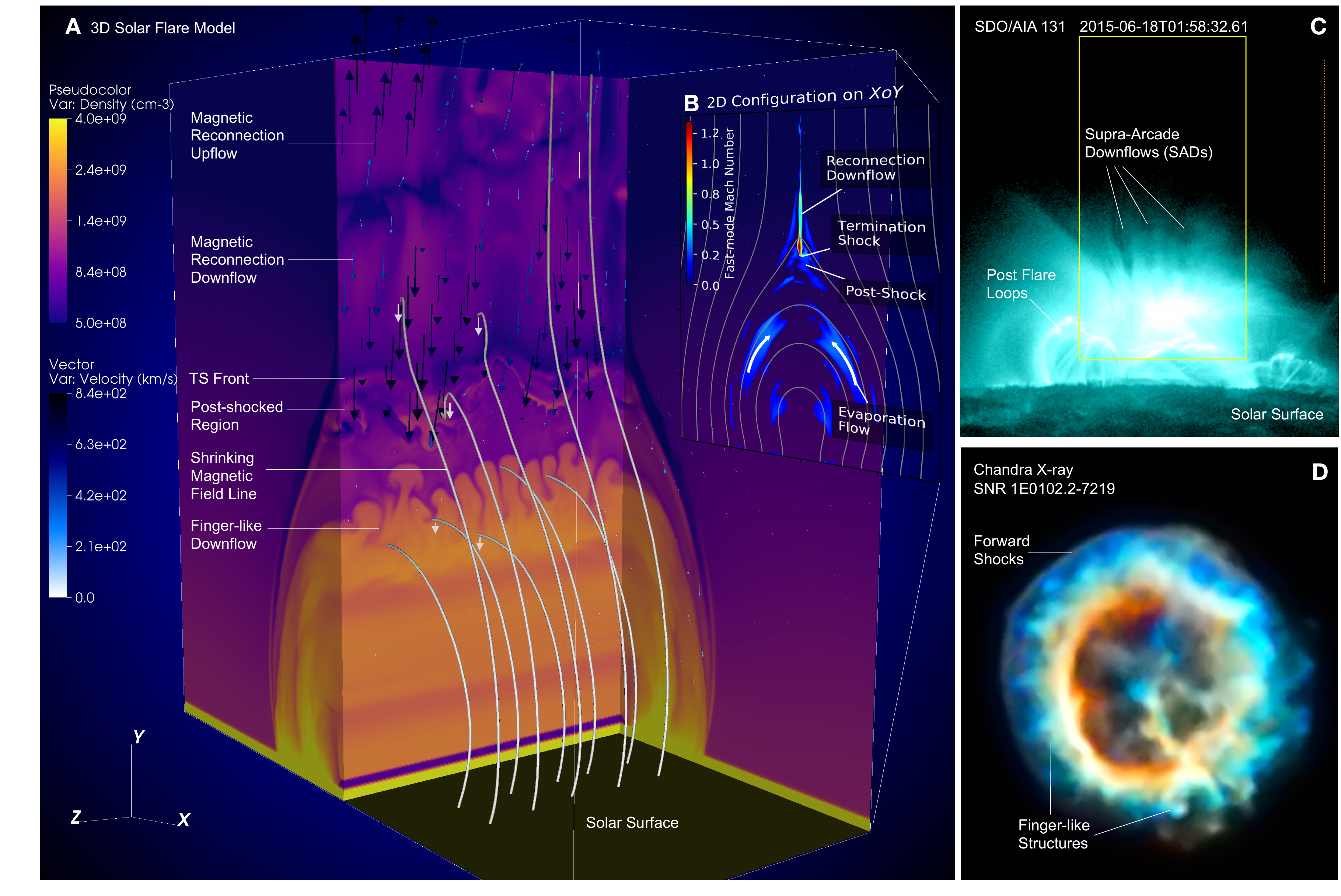}
    \caption{\textbf{Observations and 3D modeling of the energy release region of a solar flare.}
        (A) 3D modeling of the reconnection current sheet and post-reconnection flare loops at time=3.5$t_0$.
        The black/blue arrows indicate plasma flows, and solid tubes are magnetic field lines where the shrinking loop is signed using gray arrows (see animation in movie s1).
        (B) 2D slice on the $x$--$y$ plane of the 3D model in panel (A), which conforms to the standard flare model depicted in 2D.
        (C) Supra-arcade downflows (SADs) above the post-reconnection flare loops observed by SDO/AIA 131\AA\ , an EUV band covering the \ion{Fe}{xxi} spectral line at a temperature of $\sim 10^7$K\cite{O'Dwyer2010A&A...521A..21O} (see Fig. S1 and movie s2).
        (D) Finger-like structures in supernova remnants (SNRs) observed by Chandra X-ray (Credit: NASA/CXC/MIT/D. Dewey et al. \& NASA/CXC/SAO/J. DePasquale);
        }
    \label{fig:fig1}
\end{figure}

As this interface region appears different in face-on and edge-on viewing perspectives, three-dimensional (3D) numerical studies are required to reveal its true nature. 
Here, we perform 3D resistive MHD simulations with an initial standard flare  configuration by symmetrically extending a well-developed 2D flare model to 3D (Methods and Supplementary Fig. S2).
A 3D view of the flare reconnection geometry is shown in Fig.~1A.
Our model reproduces several characteristics of reconnection-driven flare phenomena in accordance with the standard flare scenario: (i) The reconnection drives fast bi-directional outflows, which can become super-magnetosonic in certain locations, inducing patchy fast-mode shocks (referred to as the ``termination shocks''); (ii) Dense flare arcades form below the interface region due to the accumulation of reconnected magnetic flux and the evaporation of plasma from the lower atmosphere (Fig.~1B); (iii) An interface region is formed below the extended reconnection current sheet and above the flare arcades, where high-speed reconnection outflows abruptly slow down. In this region, highly bent magnetic field lines initially with a cusp shape quickly relax to a more potential state with smaller curvatures (Fig.~1B).

Our model also reveals new features of the interface region not included in previous flare models: it hosts a myriad of turbulent plasma flows which, in turn, induce a mixture of strong positive and negative electric currents, where turbulent reconnection may readily occur (Fig.~2A). As the flare proceeds, multitudes of narrow, finger-like descending flows start to appear in this turbulent interface region (Fig. 1A and Supplementary movie s1).
These fingers have a lower density than the surrounding plasma. Thanks to the thermal dynamics treatment included in our simulations (Methods), we can reliably reconstruct the EUV images to compare with the observations directly. At each spatial location, the synthetic EUV intensity is obtained by integrating the density and temperature distribution of the plasma over all cells along the line of sight, folding through the instrument response (Methods). 
We show the synthetic SDO/AIA 131 \AA\ EUV map in Fig. 2(C).
The finger-like density depletion structures in the simulation appear as dark voids seen in the synthetic 131 \AA\ image, which closely resembles the observed features of SADs. We note that the density depletion nature of the SADs is supported in previous studies based on the differential emission measure inversion techniques\cite{hanneman_thermal_2014, Reeves2017ApJ...836...55R}. 

\begin{figure}[H]
    \centering
    \includegraphics[width=1.0\textwidth]{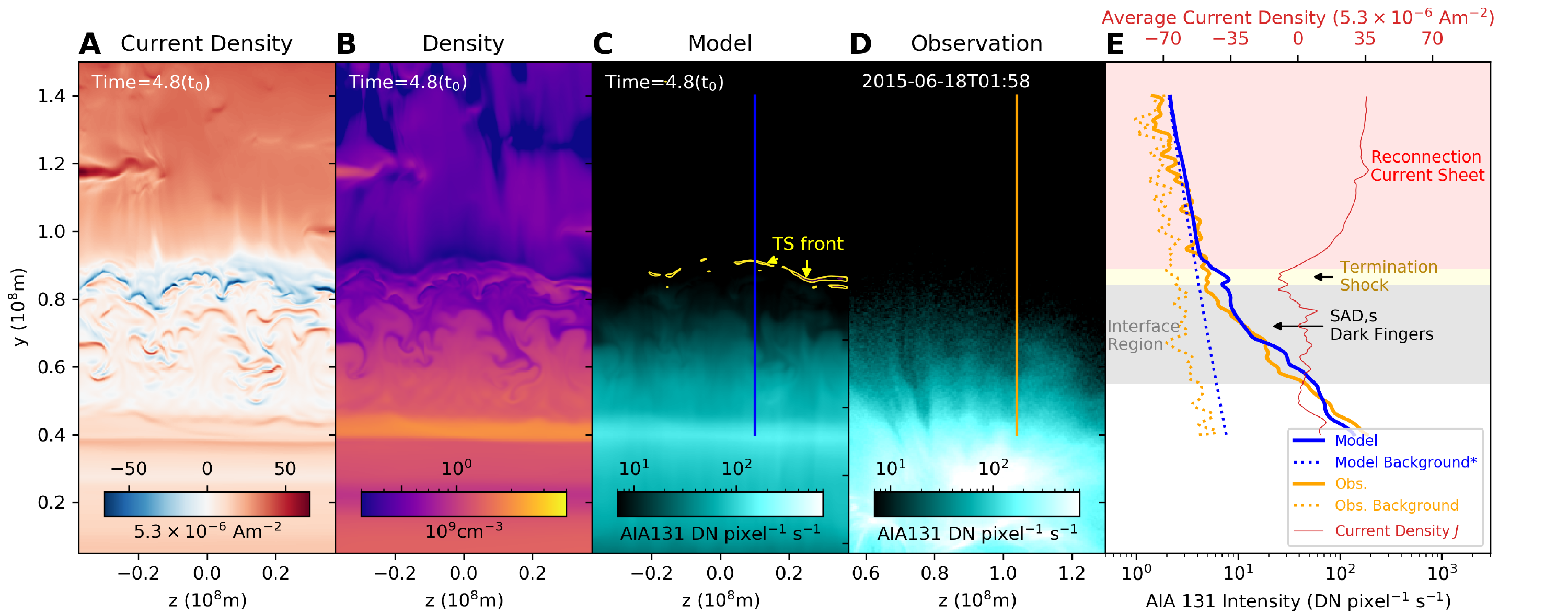}
    \caption{ \textbf{Detailed face-on view of the reconnection current sheet and the turbulent interface region in the 3D MHD model}.
    (A) The average current density ($\overline{J}$) distribution covering 20 cells near the center plane ($x$=0) showing turbulent magnetic fields in the interface region;
    (B) Plasma density at the center plane ($x$=0);
    (C) Synthetic AIA 131\AA\ image. A termination shock (TS) (identified by minimum $\nabla \cdot v$) is present below the CS; 
    (D) Observed AIA 131\AA\ intensity as same as in Fig. 1C (yellow box);
    (E) Comparison of the modeled and observed AIA 131\AA\ intensity distributions with heights. The blue and orange lines are for model prediction and observation along the sampling lines shown in panels (C) and (D). The dotted blue line shows the background corona at the beginning of the simulation, and the dotted orange line is for the observational background along the dotted sampling line as shown in Fig. 1C; 
    The red line shows the height profile of the average current density in panel (A), in which a high current density characterizes the current sheet region.
    The locations of the reconnection current sheet (red), termination shock/post-shock region (yellow), and dark finger-likes/SADs (gray) are marked.
    }
    \label{fig:fig2}
\end{figure}

Remarkably, the SADs are not located in the reconnection current sheet. Instead, they are located in the interface region below the lower tip of the current sheet. 
Such a distinction is clearly shown in Fig.~2, which demonstrates that the SADs reside in the interface region characterized by a weak, but highly non-uniform current density as well as a steeply decreasing intensity profile in height. In contrast, the reconnection current sheet, by definition, features a strong current density and shows a much shallower intensity profile. The two regions are demarcated by the lower end of the reconnection current sheet where a fast-mode shock may be present. Such distinctively different intensity profiles between the interface region and the current sheet region are also evident in observations (solid orange curve in Fig.~2E). This distinction is due to the different variations in plasma density and column depth as a function of height between the two regions when observed from a face-on viewing perspective of the current sheet.
In both the observed and simulated intensity profiles, the SADs are located in the region below the lower tip of the current sheet, which has a much steeply decreasing intensity profile (Fig. 2E).

Consistent with the observations of SADs, the propagation speeds of the finger-like structures in our simulations are much slower than that of the Alfv\'{e}nic reconnection outflows. In Fig. 3, we trace the downward plasma flows along a vertical cut in the simulation, producing time-distance maps (right panels). Two types of downflow features can be clearly identified: 
(i) fast flows with a speed of $\sim$400--600 km s$^{-1}$
that are either dense or tenuous than surrounding plasma inside the current sheet region, and (ii) slower flows with a speed of $\sim$50--200 km s$^{-1}$ below the lower end of the current sheet that are less dense than their surroundings.
The former, which have slightly sub-Alfv\'{e}nic
(the local Alfv\'{e}n speed is $\sim$750--900 km s$^{-1}$ in the inflow region just outside the reconnection current sheet)
, are consistent with the properties of reconnection outflows predicted in MHD and PIC simulations\cite{2018PhPl...25j2120H}. Observationally, they may be the counterpart of the fast bright downflows observed in the post-eruption plasma sheets\cite{Liu2013ApJ...767..168L}.
The latter, in both strong and weak reconnection conditions (Methods, Fig. S4),  have speeds that are consistent with the speeds of the SADs reported in the literature\cite{Savage2012ApJ...747L..40S}.

\begin{figure}[H]
    \centering
    \includegraphics[height=0.9\textwidth]{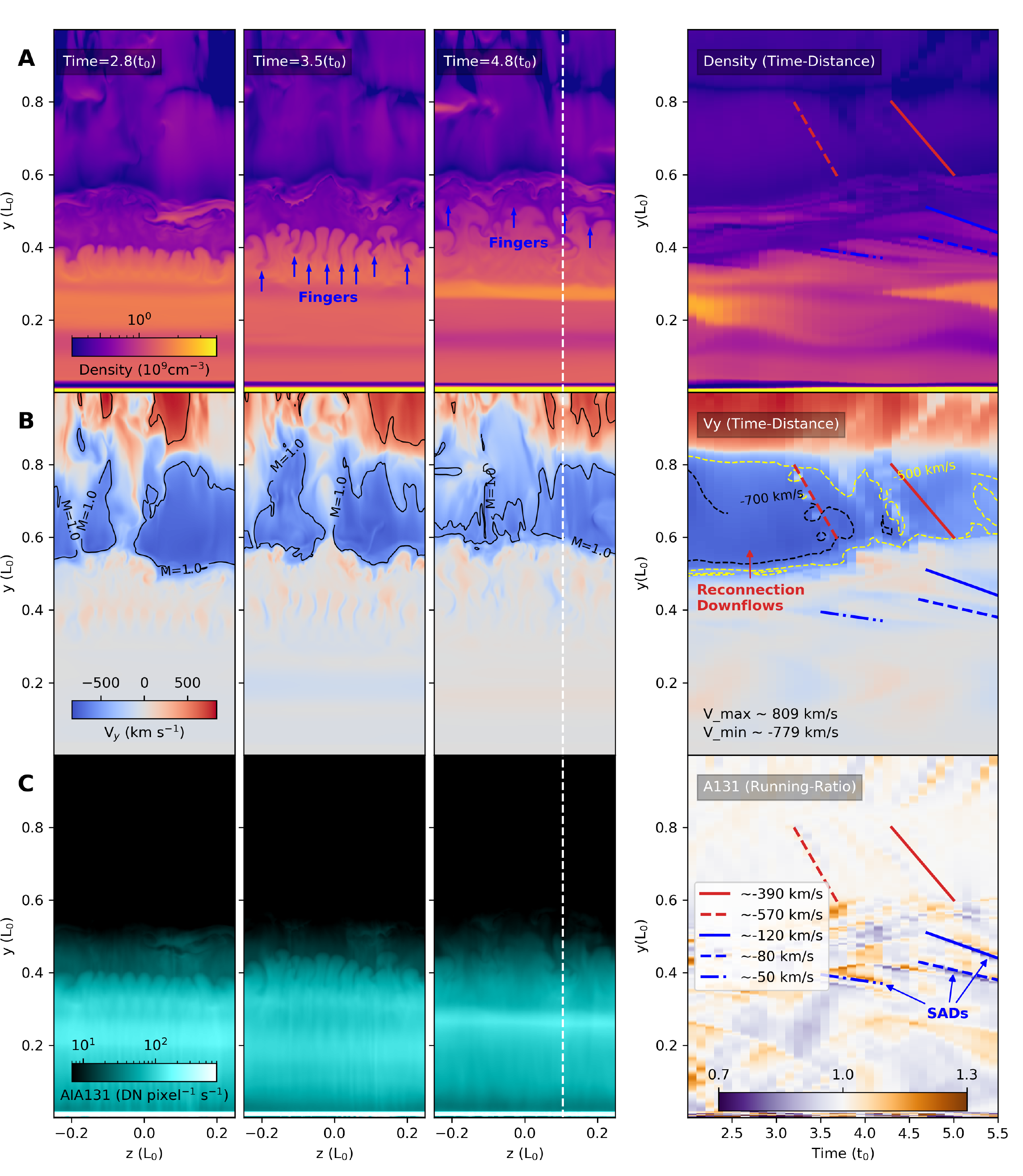}
    \caption{
    \textbf{Development of finger-like dynamic structures in the interface region underneath the current sheet.} 
    (A, B) Plasma density and velocity component ($V_y$) at the central $y$-$z$ plane in the 3D MHD model with the current sheet viewed face on. The finger-like structures are signed by blue arrows. The black solid contour lines on the $V_y$ maps enclose regions where the fast-mode magnetosonic Mach number $M_A$ exceeds unity.
    (C) Synthetic AIA 131\AA\ (and 94\AA\ in Fig. S3) EUV maps. The right panel of each row shows the space-time evolution of the respective physical properties/emission intensity along the vertical dashed line (at $z$=0.1), which cuts through both the reconnection current sheet and the finger-like structures. 
    Various plasma flows can be seen as coherent features with a distinctive slope. Examples of the fast (390--570 km~s$^{-1}$) reconnection outflows in the current sheet and the relatively slow (50--120 km~s$^{-1}$) finger-like SADs beneath the current sheet are outlined as red and blue slopes, respectively.
    }
    \label{fig:fig3}
\end{figure}

Due to the distinctively different locations and speeds between the SADs and the reconnection outflows, in stark contrast to previous interpretations,
we conclude that the SADs are indirect results of reconnection outflows, rather than the outflows themselves.
As we demonstrate below, they are self-organized structures formed in the turbulent interface region due to a mixture of the Rayleigh-Taylor instability (RTI) and the Richtmyer-Meshkov instability (RMI), similar to the case of the post-shock region in supernova remnants\cite{Miles2009} (see Fig. 1D).

The RTI/RMI usually develops at the material interface separated by different densities due to vorticity deposited by gravity or transmitted shocks. 
In our model, the initial development of perturbations can be understood as the result of RTI/RMI in the weak shock limit, where the plasma flow interacts with the flare arcade (Supplementary Figs. S5--S7).
The instabilities can appear at both the upper and bottom boundary of the interface region (Fig. 4A): (i) the lower one separating dense inner post-flare loops and the upper tenuous region containing newly reconnected shrinking magnetic field, and (ii) the higher density boundary due to compressed post-shocked plasma below the current sheet.
These instabilities cause the rippled appearance of the interfaces and the formation of ``bubbles'' and ``spikes'', and eventually the formation of dark fingers with reduced density in the non-linear phase of the RTI/RMI (Fig. 4B).
Moreover, the RTI/RMI repeats at the top surface of the fully developed ``bubbles'' where more new ``spikes'' appear continuously (Fig. 4C) and, consequently, results in the development of a turbulent ``interface'' region. This scenario is consistent with the intermittent features observed in SADs\cite{savage_quantitative_2011}.

\begin{figure}[H]
    \centering
    \includegraphics[width=1.0\textwidth]{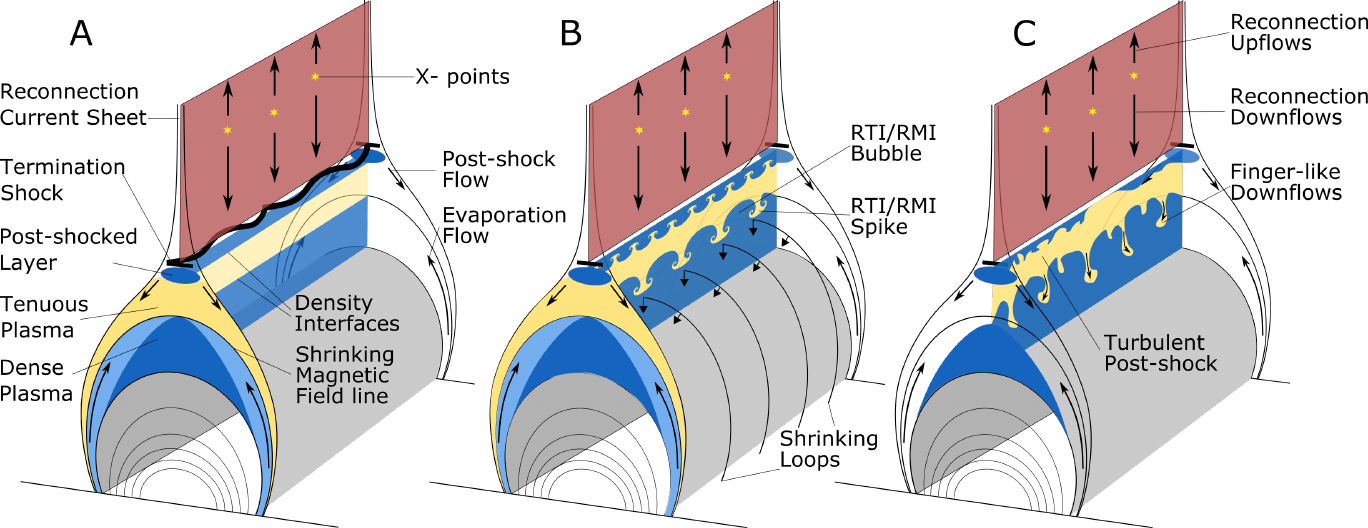}
    \caption{\textbf{Schematic diagram of the development of RTI/RMI instabilities in the turbulent interface region beneath the reconnection current sheet}. 
    (A) Schematic of the 3D configuration of the flare reconnection geometry as shown in Fig. 1(A).
    (B) Instabilities develop in turbulent interface region, causing ``spikes'' and ``bubbles'' characteristic of the RTI/RMI scenario. 
    (C) The downwards ``spikes'' evolve into under-dense finger-like flow structures, as observed in the observations, during the non-linear phases of the instabilities when secondary instabilities also play a role.
    }
    \label{fig:fig4}
\end{figure}

In addition to the SADs, our model also offers insights for interpreting other highly relevant observed phenomena. For example, Fig. 1(A) (and the accompanying animation) shows both fast-contracting magnetic loops in the reconnection outflow region and relatively slower ones in the underlying interface region, which may exhibit themselves in EUV/SXR observations as bright plasma downflows or shrinking loops  \cite{savage_quantitative_2011,Savage2012ApJ...747L..40S,Liu2013ApJ...767..168L}. In addition, with sufficient contrast against the background, the relatively faint reconnection outflows in the current sheet region (Figs. 3(B) and (C)) may also be detected. Although rather rare in the literature, there have been reports of such oppositely directed plasma outflows above the flare arcade, which are sometimes referred to as ``disconnection events''
\cite{Savage2010ApJ...722..329S,2020ApJ...900...17Y}.
Further joint observation--modeling studies are required to place these phenomena observed in different flare events from various viewing perspectives into a coherent, reconnection-driven physical picture.

Finally, our results reveal the rich complexity of the highly turbulent interface region, which may hold the key to magnetic energy release and conversion during solar flares. As discussed above, firstly, the structure and dynamics of the plasma developed in the interface region offer new insights into interpreting multiple observed phenomena, including SADs and other reconnection-driven plasma outflows.   
Another important implication is on the derivation of the dimensionless reconnection rate ($M = v_{\rm in}/v_{A}$), which is an important measure of the reconnection efficiency defined as the ratio between the speed of the inflowing plasma toward the reconnect current sheet and the local Alfv\'{e}n speed (Methods). Previously, the observed speeds of plasma flows above the flare arcade, including SADs, are often used as a proxy for $v_{A}$\cite{Savage2012ApJ...747L..40S,Su2013NatPh...9..489S}. However, in some cases, this practice may lead to a significant overestimate of the reconnection rate, because certain observed plasma outflows are substantially slower than the Alfv\'{e}n speed as suggested by our model.
Last but not least, the turbulent flows, and fast- and slow-mode shocks in the interface region can also lead to plasma compression and heating, while the corresponding complex magnetic configuration and the turbulent plasma can facilitate efficient particle acceleration and small-scale magnetic reconnection\cite{2020NatAs...4.1140C,2020Sci...367..278F}. 
Our results highlight the importance of including detailed 3D effects in studying magnetic reconnection and interpreting certain key observable phenomena associated with flare energy release.

\clearpage
\section*{Methods}
{\bf Numerical Simulation Code:}
The 2.5D/3D magnetohydrodynamics (MHD) simulations are performed using the open-source astrophysical code: {\it Athena}\cite{2008StoneApJS..178..137S} (https://github.com/athena-team/athena). {\it Athena} solves the full equations of MHD in one, two, or three dimensions. The system of equations evolved by {\it Athena} includes Ohmic resistivity, ambipolar diffusion (for partially ionized plasmas), the Hall effect, isotropic and anisotropic thermal conduction, optically thin radiative cooling, and gravity (either self-gravity or static gravitational potential). The code is based on the directionally unsplit high-order Godunov method, and it combines the corner transport upwind (CTU) and constrained transport (CT) methods. It provides superior performance for capturing shocks as well as contact and rotational discontinuities. The low numerical diffusion provides a significant advantage for resistive MHD simulations of processes like magnetic reconnection. Another feature of Athena is that the magnetic divergence-free constraint is satisfied to very high accuracy ($10^{-10} \sim 10^{-12}$), which is necessary for performing magnetic reconnection simulations.

\noindent{\bf Observational Data:}
We analyze images observed by the Atmospheric Imaging Assembly (AIA) on board the Solar Dynamics Observatory (SDO)\cite{2012lemenSoPh..275...17L}. SDO/AIA is a narrowband imaging instrument in extreme ultraviolet (EUV) and UV wavelengths. 
SDO/AIA has the ability to capture dynamic phenomena with a high temporal and spatial resolution (cadence of 12 s and 0.6'' pixel$^{-1}$). 
It includes six wavebands that measure emission predominately from coronal iron lines including 94 \AA, 131 \AA, 171 \AA, 193 \AA, 211 \AA, and 335 \AA. 
The temperature sensitivities range from $\approx$~$4 \times 10^5$ K (\ion{Fe}{viii}) to $\approx$~2 $\times$ $10^7$ K (\ion{Fe}{xxiv}). Flare fans and SADs have been observed in all hot AIA channels (including 94, 131, 193, 335 \AA), but are best seen in the 94 and 131 {\AA} images. 
Therefore, in this work, we focus on the 94 {\AA} and 131 {\AA} bands which are, respectively, mainly sensitive to \ion{Fe}{xviii} plasma, formed at temperatures of $\sim$~6 MK, and \ion{Fe}{xxi}, formed at temperatures greater than 10 MK in solar flare regions\cite{O'Dwyer2010A&A...521A..21O}. 
The AIA images are obtained and calibrated using the standard calibration routines in AIApy (https://pypi.org/project/aiapy/) and SunPy (https://github.com/sunpy).
In this work, we show the flare fan and SADs features of well-reported eruptive flare event observed by AIA on 2015-06-18, as shown in Fig. 1C and movies s2.

\noindent{\bf Simulation Setup:}
We perform a series of resistive magnetohydrodynamic (MHD) simulations to investigate the dynamic features of magnetic reconnection current sheets and the plasma flows above flare loops during flare eruptions.   The evolution of the system can be obtained by solving the initial and boundary value problem governed by the resistive MHD equations. 
We then solve the following MHD equations using the Athena code:
\begin{eqnarray}
\frac{\partial\rho}{\partial t} & + & \nabla\cdot(\rho \mathbf{v})=0, \label{eq:mass} \\
\frac{\partial\rho\mathbf{v}}{\partial t} & + & \nabla\cdot(\rho\mathbf{v}\mathbf{v}-\mathbf{B}\mathbf{B}+\mathbf{P^{*}})=-\rho \nabla \phi, \label{eq:momentum} \\
\frac{\partial \mathbf{B}}{\partial t} & - & \nabla\times( \mathbf{v}\times \mathbf{B})=\eta_{m}\nabla^{2} \mathbf{B}, \label{eq:induction} \\
\frac{\partial E}{\partial t} & + & \nabla\cdot[(E+P^{*})\mathbf{v} - \mathbf{B}(\mathbf{B}\cdot\mathbf{v})]=S, \label{eq:energy}
\end{eqnarray}
where $\rho$, $\mathbf{v}$, $\mathbf{B}$, $E$ are plasma density, velocity, magnetic field, and total energy density, respectively. $\phi$ is static gravitational potential, ${\mathbf{P^{*}}}$ is a diagonal tensor with components $P^{*} = P + B^{2}/2$ (with P being the gas pressure), and $E$ is the total energy density, given by
\begin{equation}
E=\frac{P}{\gamma-1} + \frac{1}{2}\rho v^2 + \frac{B^2}{2}, 
\end{equation}
where ${\gamma = 5/3}$ is the adiabatic index,
and the energy source term $S = \mu_0 \eta_{m}{j^2} + \nabla_{||}\cdot\kappa\nabla_{||} T + H_{cooling,heating}$, which includes Ohmic dissipation, thermal conduction, radiative cooling and coronal heating terms. Here $\mu_0, \eta_m$ and $\kappa$ are the magnetic permeability of free space, magnetic diffusivity, and parallel component of the Spitzer thermal conduction tensor, respectively.

The above equations are solved in non-dimensional forms in the simulation, and can then be scaled to physical units based on observational values during solar eruptions. In a typical two-ribbon solar flare, the characteristic values can be chosen as the following: $L_0=1.5 \times 10^8$\ m, $B_0=0.00099$\ Tesla, $\rho_0=2.5 \times 10^{14}$\ m$^{-3}$, $T_0=1.13 \times 10^8$\ K, $V_0=1.366 \times 10^6$\ m/s, and $t_0=109.8$\ s. 
The simulations are performed on Cartesian coordinates with uniform cells in the domain ([-0.5, 0.5], [0, 1.0] ,[-0.25, 0.25]) in the X-, Y-, and Z-direction, respectively. The finest grid size and typical time step are 0.00173 and about 0.0001 in normalized units, respectively. These correspond to $\sim$260 km and $\sim$0.01 s in physical units based on the above characteristic parameters. The primary simulation parameters are also listed in Table 1.
The boundary conditions at the bottom boundary are set to ensure that the magnetic field is line-tied to the photosphere, and the plasma does not slip as well. The other boundaries of the simulation domain are all open, such that the plasma and the magnetic flux are allowed to enter or exit freely through them \cite{shen_numerical_2011,Shen_2018}.

Anisotropic thermal conduction, static gravity, and radiative cooling terms are included in this model. We use the super timing-step scheme\cite{Meyer2012MNRAS.422.2102M} to solve the thermal conduction part in Equation (\ref{eq:energy}) and include the optically thin cooling term ($n_e n_H Q(T)$). Here, $Q(T)$ is calculated using a piece-wise linear approximation \cite{Klimchuk2008ApJ...682.1351K}. 
An additional density-dependent coronal heating term ($\rho H_0$) is included in the energy equation to balance the radiative cooling in the background coronal regions.
To include the contribution from the dense chromosphere, we set up a thin cooler and denser layer at the bottom boundary to represent the chromosphere and transition region following the practice of other successful flare models \cite{yokoyama_magnetohydrodynamic_2001,takasao_magnetohydrodynamic_2015}. The minimum temperature at the bottom boundary is lower than $5.5 \times 10^3$ K. The transition width from coronal temperatures to the bottom of the simulation domain is about 2500 km according to general estimations for the height of the chromospheric layer.  At the bottom of the simulation domain, where the plasma density is extremely high, we assume that the cooling terms are always balanced by coronal heating by adding an artificial heating term. However, we note that there are additional contributions from the chromosphere due to the precipitated-electron-induced chromospheric evaporation, which are not modeled in our MHD simulations.

Our simulations are comprised of 2.5D models and 3D models. 
The 2.5D model is run first to form the classic Kopp-Pneuman configuration of two-ribbon flares\cite{kopp_magnetic_1976}.
We initialize the system from a pre-existing Harris-type current sheet along the $y$-direction with a non-dimensional width $w=0.03$. The current sheet is in mechanical and thermal equilibrium and separates two magnetic field regions with opposite polarities (see details in \cite{shen_numerical_2011,Shen_2018}). The only difference with our previous setting is that we include the guide field ($B_z$ component) inside this initial current sheet to balance the gas pressure. 
At the beginning of the 2.5D simulations, we introduce a perturbation on the initial Harris-type current sheet following the previous models\cite{Shen_2018}. The magnetic field then starts to diffuse at the perturbation position,  where the two sides of the current sheet slowly move towards each other due to the Lorentz force attraction.
A pair of reconnection outflows gradually form, and closed magnetic field loops appear at the solar surface due to the accumulation of reconnected magnetic flux. 
Fig. S2 shows the above evolution process for current density $J_z$, velocity component $v_y$, and gas pressure $p$.

Once closed flare loops are well-formed at the bottom of the simulation domain, we start three-dimensional simulations at Time$=17, 18(t_0)$ for Case A and B, respectively. The magnetic configurations from the 2.5D simulations are used as the initial conditions.
The system can self-consistently evolve by symmetrically extending all primary variables from the 2D plane ($xoy$) to the third axis ($z$-) and using the same time-step as in the 2.5D simulations in the full 3D framework. 
We also use the same boundary conditions and keep all parameters, including Reynolds number, cooling rates, and gravity, consistent with the 2.5D simulation.

\noindent{\bf Synthetic SDO/AIA Intensity:}
To compare our models with observations, we have calculated synthetic SDO/AIA intensities in two channels (94 \AA\ and 131 \AA), which are sensitive to high-temperature flare plasma\cite{Boerner2012SoPh..275...41B}. Once we compute the plasma density and temperature on each cell from the 3D MHD simulations, the intensity (count rates) of AIA filter-band $m$ is obtained by using the formula \[I_m = \sum_{i} n_{e, i}^2 f(T_{e, i}) dl\] summed over all cells along the line of sight (LOS). Here, $n_e$ is the plasma number density, $T_e$ is the temperature, and $dl$ is the column depth of each grid $i$ along the LOS. $f(T_e)$ is the SDO/AIA response function of AIA filter-band $m$, which are obtained using the CHIANTI v.9.0.1 database\cite{Dere_etal_2019ApJS..241...22D} and the ``aia\_get\_response.pro'' routine in SolarSoft (SSW)\footnote{(https://www.lmsal.com/solarsoft/ssw)}. We assume equilibrium ionization and coronal abundances\cite{Feldman1992PhyS...46..202F}. 

\noindent{\bf Strong/Weak Magnetic Reconnection Situation:}
Following two classic solar flare systems formed in supermagnetosonic and submagnetosonic reconnection regimes \cite{Forbes1996ApJ...459..330F}, we investigated the magnetic-reconnection-related plasma flows in two cases with fast and slow magnetic reconnection, respectively.
The mechanism for driving fast magnetic reconnection in solar flares is an ongoing research topic. Several reconnection models have been suggested, such as a Sweet-Parker current sheet with anomalous resistivity \cite{Priest2000mrmt.conf.....P}, Petschek-type reconnection\cite{Petschek1964NASSP..50..425P,Forbes_Priest1987RvGeo..25.1583F,Yokoyama1994ApJ...436L.197Y}, turbulent reconnection \cite{Strauss1988, Lazarian2012SSRv..173..557L, Lazarian2020PhPl...27a2305L}, and plasmoid instabilities that develop inside a reconnecting current sheet\cite{Loureiro2007PhPl,Bhattacharjee2009PhPl...16k2102B, Ni2010PhPl...17e2109N}. 
In addition, it has also been suggested that the different dissipation processes may work together to achieve a more efficient diffusion\cite{Mei2012MNRAS.425.2824M,Ye2019MNRAS.482..588Y}.  Based on our previous numerical experiments\cite{shen_numerical_2011, Shen_2018}, we set a uniform resistivity throughout the whole simulation domain, which gives a constant magnetic Reynolds number (${R_m \sim 5 \times 10^4}$) in Cases A to drive the fast magnetic reconnection. 
The magnetic reconnection becomes bursty once fine structures (e.g., magnetic islands due to tearing instabilities) appear inside the reconnecting current sheet. Therefore, the reconnection outflows can exceed the local fast-mode magnetosonic speed that drives the termination shocks at the lower tips of the reconnecting current sheet\cite{Shen_2018}.

As a comparison, we run Case B with a low magnetic Reynolds number (${R_m} \sim 10^4$) which just approaches the lower threshold for triggering the plasmoid instability in the 2D framework\cite{Ji2011PhPl...18k1207J}. The equivalent diffusion in MHD simulations usually is slightly larger than the input physical resistivity due to the numerical diffusion (see details in \cite{shen_numerical_2011}). In Case B, the magnetic reconnection gradually takes place over a longer period of time without the development of magnetic islands, causing relatively weaker magnetic reconnection outflows in the 2.5D models. Therefore, Case A is used to simulate the unsteady magnetic reconnection process with predominantly supermagnetosonic outflows and, consequently, the formation of termination shocks, while Case B reveals the relatively weaker reconnection in the submagnetosonic regime under most circumstances. We estimate the non-dimensional reconnection rate ($M = v_{\rm in}/v_{A}$) \cite{Cassak2017JPlPh..83e7101C} in Case A and Case B by monitoring the ratio of reconnection inflow speed and the local Alfv\'enic speed near the $x$- points on a set of planes along the $z$-axis. During the period we are interested in, the non-dimensional median reconnection rate in Case A is larger than that of Case B by about a factor of 2, and it ranges from $\sim$0.05 to 0.08 in Case A and $\sim$0.02 to 0.05 in Case B, respectively.
We notice that the inhomogeneous evolution in 3D reconnecting current sheets may enhance the local reconnection rate even in Case B, which is not strongly limited by the Reynolds number \cite{Huang_2016ApJ...818...20H, Yang2020ApJ...901L..22Y}. Therefore, the outflows in Case B can occasionally reach supermagnetosonic speeds at certain locations/times, even though the overall reconnection process is relatively slow. In Fig. S4, we show such a case where a localized region with supermagnetosonic speeds can be identified (solid contour in panel (B)). 
The primary variables, density and velocity, and integrated synthetic AIA 94 and AIA 131 intensity are shown in Fig. S4 in a similar form to those shown in Fig. 3. Particularly, several dark fingers (or SADs) can be seen from both the plasma density maps and the synthetic AIA 131 \AA\ and 94 \AA\ images. The downflow speeds ($\sim$ 60 km/s) can also be obtained from AIA running-ratio images, which are within the same range as those estimated from Fig. 3. Our practice demonstrates that the formation of the under-dense, finger-like structures in the interface region beneath the reconnection current sheet may be a universal phenomenon under very different reconnection regimes.

\noindent{\bf RTI/RMI at The Density Interface:}
We performed a detailed analysis of the initial development of the perturbations by identifying the variation of several primary physical variables crossing the density interface. Figure S5 shows the evolution of the density interface at the central plane ($x=0$), as shown in Fig. 3. At time 1.8$t_0$, a density interface appears around $y = 0.355L_0$ separating the dense plasma ($\rho > 6 \rho_0$) from the tenuous gas ($\rho \sim 2\rho_0$) as shown in Fig. S5(A).
At later times (t=1.9-2.0$t_0$), this density interface is disturbed and develops ripples due to the linear RTI/RMI.
We show the gas pressure and vertical velocity component $V_y$ (Figs. S5(E)--(G)) as well as their respective gradients (Figs. S5(H)--(J)) at three selected times (t=1.8, 1.9, and 2.0$t_0$) along a sample vertical dashed line.
The location of the density interface can be identified as the sharp transition in the gas pressure, which corresponds to the local minimum in the gradient profiles, as shown by the gray shadow regions in Figs. S5(H) and (I). 
Both density and pressure gradients are negative, which means that they point downward across the interface, while the vertical velocity gradient (or $\nabla V_y$) is directed upwards, as shown in Fig. S5(J). 

The property of this interface matches well the condition for driving the classic RMI. 
In general, the RMI occurs when the sign of the density and pressure gradients are opposite to each other ($\nabla \rho \cdot \nabla p < 0$) if the shock propagates from the light to dense gas\cite{Zhou2017PhR...720....1Z}. Alternatively, if the shock directs from the heavy to light gas, the RMI occurs when $\nabla \rho \cdot \nabla p > 0$.
In Fig. S5(J), an upward velocity gradient bump can be clearly seen around $y=0.35 L_0$ in the dense plasma side just below the interface at time 1.8(t$_0$).
In this case, the pressure and density gradient follows the second case above, where the shock travels from dense to tenuous gas with ${\nabla \rho \cdot \nabla p > 0}$.
The random numerical perturbations at this density interface then can develop into ripple structures and eventually evolve into ``spikes'' and ``bubbles'' at the two sides of the interface. 
At later times (red line in Fig. S5(J)), this upward $\nabla v_y$ bump developed two separated components: one fast upward part on the upper side and a slower downward one on the downside that is similar to the classic RM instability.

Further, we investigate the evolution of the interface during the non-linear RMI phases and compare it with a theoretical model. In Fig. S6(A), we show the evolution of bubbles at the density interface in Case A in which the different colors indicate different times in the simulation ranging from 2.0 -- 2.6$t_0$. In general, the RMI quickly evolves into non-linear phases once the penetration depth of the bubble (or spike) reaches a significant fraction of the perturbation wavelength\cite{Alon1995PhRvL..74..534A}, as can be seen in Fig. S6. We then monitor the density variation across the interface and obtain the average Atwood number ($A=(\rho_d - \rho_t)/(\rho_d + \rho_t)$: a measure of the relative density jump), in Fig. S6B. It is clear that this interface follows the low Atwood number condition where $A$ ranges from $\sim$0.5 to 0.44 during this period.  Theoretical models of the RMI and relevant simulations suggest that a power-law function can be used for approximately describing the growth of the heights of the bubbles (or spikes) in the case of multi-mode initial perturbations:
\begin{equation}
    h(t) \propto \tau(t)^{\theta},    
\end{equation}
where $h$ is the amplitude of perturbation, $\tau(t)$ is a linear function of time $t$, and $\theta$ is the power-law growth index. 
This power law has been widely investigated based on theoretical models, experiments, and simulations where $\theta$ has been suggested to range from $\sim $0.25 to 2/3\cite{Zhou2017PhR...720....1Z}. For instance, Alon et al.\cite{Alon1995PhRvL..74..534A} predicted that $\theta \sim 0.4$ for both bubble and spike for all values of $A$. In Fig. S6(C), we plot the predicted growth profile of the bubbles based on Alon's model by setting the initial perturbation wavelength and velocity to 0.02 and comparing the simulation profiles with the prediction. Figure S6(C) shows that the two profiles roughly match. We conclude that the non-linear development of the bubble agrees with the classic RMI theory, although the power-law index may slightly differ in some simulations. Because the Atwood number is less than 1 in our cases, the growth of the bubble and spikes is not symmetrical in the later phases\cite{Alon1995PhRvL..74..534A}. We will briefly discuss the behavior of the spikes in the following sections.
We note that, occasionally, upward plasma motion in the interface region can also be identified. Indeed, observations of such phenomena have been reported in the literature \cite{Mckenzie2013ApJ...766...39M, Samanta2021Innov...200083S}. However, these may only represent localized events in the interface region where the overall downward-contracting magnetic loops dominate the plasma motion.

Although the discussions above focus on the RMI, in practice, the RTI and the RMI might be involved simultaneously in various environments\cite{Aschenbach1995Natur.373..587A,Balick2002ARA&A..40..439B,Attal2015ShWav..25..307A}. The growth rate of instabilities is then different from the pure RMI system, especially in situations with gravity including in our MHD models. Therefore, in our case, it is reasonable to consider the initial development of the instabilities as the combined result of both RTI and RMI\cite{Chen2018PhFl...30j2105C}.

We note that there is a significant difference between our case and the classic hydrodynamic RTI/RMI because of the presence of a magnetic field\cite{wheatley_magnetohydrodynamic_2015}. It is beyond the scope of the current work to directly compare our simulations with other RTI/RMI experiments that include the magnetic field, because the magnetic field configuration can be different on a case-by-case basis and the magnetic field also dynamically evolves during a flare eruption.
However, for demonstration purposes, we perform a qualitative analysis to discuss the effects of the magnetic field by monitoring the variation of plasma $\beta$, the ratio of the plasma gas pressure to the magnetic pressure. Fig. S5(E) shows the average $\beta$ \ in the  y-direction at these times. It is worth noting that the RTI/RMI features appear at the interface region ($\beta \approx 1$), where the plasma conditions change from magnetically dominated ($\beta < 1$) in the upstream region and elsewhere in the flaring site to the fluid dominated ($\beta > 1$) regime.

A similar RTI/RMI feature can be found at the upper location just below the lower tip of the current sheet, as shown in Fig. S7. 
The upper dense layer forms due to the compression of the post-shocked plasma during the intermittent reconnection, which is commonly predicted in solar eruption models.
The profiles of the primary values across this density interface are similar to those in Fig. S5, except that the density structure is reversed.

\subsection*{Data Availability}
The SDO/AIA data are publicly available and obtained using the Sunpy module \emph{Fido}.

\subsection*{Code Availability} 
The MHD code is achievable on (https://princetonuniversity.github.io/Athena-Cversion/).
The AIA data are analyzed using the Sunpy (https://github.com/sunpy), and AIApy packages\\ (https://pypi.org/project/aiapy). 
The SolarSoft (SSW) package is obtained from \\(https://www.lmsal.com/solarsoft/ssw). The Chianti atomic data are obtained through\\ (https://www.chiantidatabase.org/).

\newpage
\section*{Acknowledgements}
\noindent
The authors thank L. Guo for the help on the modeling setup, and thank J. Raymond, N. Murphy and J. Lin for helpful discussions.
The AIA is an instrument on SDO, a mission of NASA. CHIANTI is a collaborative project involving George Mason University, the University of Michigan (USA), and the University of Cambridge (UK). The computations in this paper were conducted on the Smithsonian High Performance Cluster (SI/HPC), Smithsonian Institution (https://doi.org/10.25572/SIHPC).
C. S. and K. R. are supported by NSF grants AST-1735525, AGS-1723313, AGS-1723425 to SAO. B. C. and S. Y. are supported by NSF grants AGS-1654382, AGS-1723436, and AST-1735405 to NJIT. V. P. acknowledges support from NSF SHINE grant AGS-1723409. X. X. is supported by the Chinese Academy of Sciences (CAS) grants XDA17040507 and QYZDJ-SSWSLH012, NSFC grant 11933009, Yunnan Province grants 2018HC023, and  the scholar ship granted by the China Scholarship Council (CSC) under file No. 201904910573.

\section*{Author Contributions}
C. S. performed the MHD simulations and analyzed the results. B. C. and K. R. proposed the study and contributed to the modeling setup and results analysis. S. Y., V. P., and X. X. contributed to EUV data collection, analysis, and visualization. C. S. and B. C. led the manuscript writing, and all authors discussed the results and commented on the manuscript.

\newpage
\begin{table}[h!]
\centering
\begin{tabular}{ |p{2cm}||p{4cm}|p{4cm}|p{2.5cm}|  }
 \hline
 \multicolumn{4}{|c|}{Parameters and Setup in MHD simulations} \\
 \hline
 No. & Grids &Magnetic Reynolds Number R$_m$&Time for Starting 3D (t$_0$)\\
 \hline
 Case A   & $579\times576\times288$    & 5 $\times$ 10$^4$ &   17\\
 Case A$_{2.5D}$   & $579\times576$    & 5 $\times$ 10$^4$ &   -\\
 Case B   & $575\times576\times288$    & 10$^4$            &   18\\
 Case B$_{2.5D}$   & $575\times576$    & 10$^4$ &   -\\
 \hline
\end{tabular}
\caption{Primary parameters for different cases. }
\end{table}

\newpage
\includegraphics[width=0.8\textwidth]{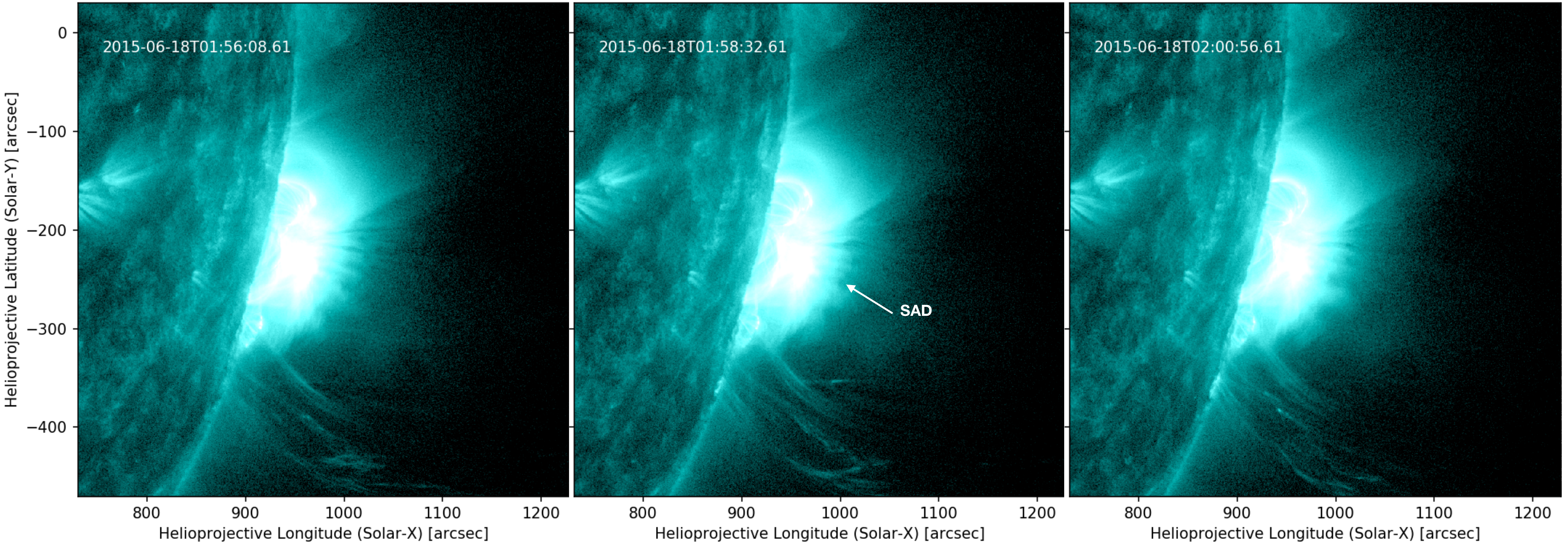}

\noindent {\bf Fig. S1.} \textbf{The example of solar eruptive flare events featuring supra-arcade downflows (SADs).} This event is observed by SDO/AIA 131 \AA\ on 2015 June 28 (see the animation in movie s2).

\newpage
\includegraphics[width=0.8\textwidth]{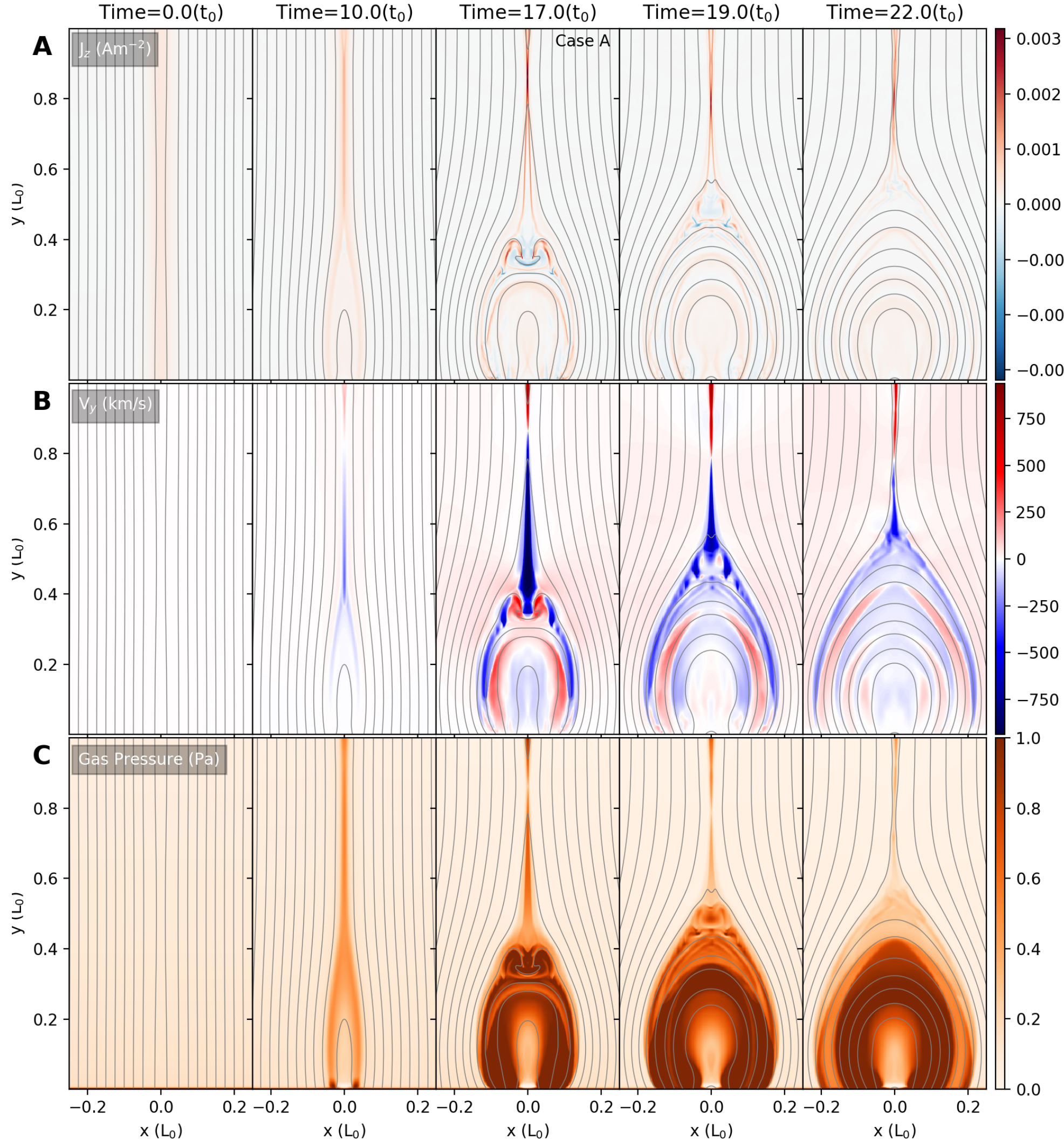}

\noindent {\bf Fig. S2.} \textbf{Evolution of current-density ($J_z$), velocity component ($V_y$), and gas pressure ($P$) in 2.5D simulations for Cases A}. The solid curves show the magnetic field lines.

\newpage
\includegraphics[width=0.8\textwidth]{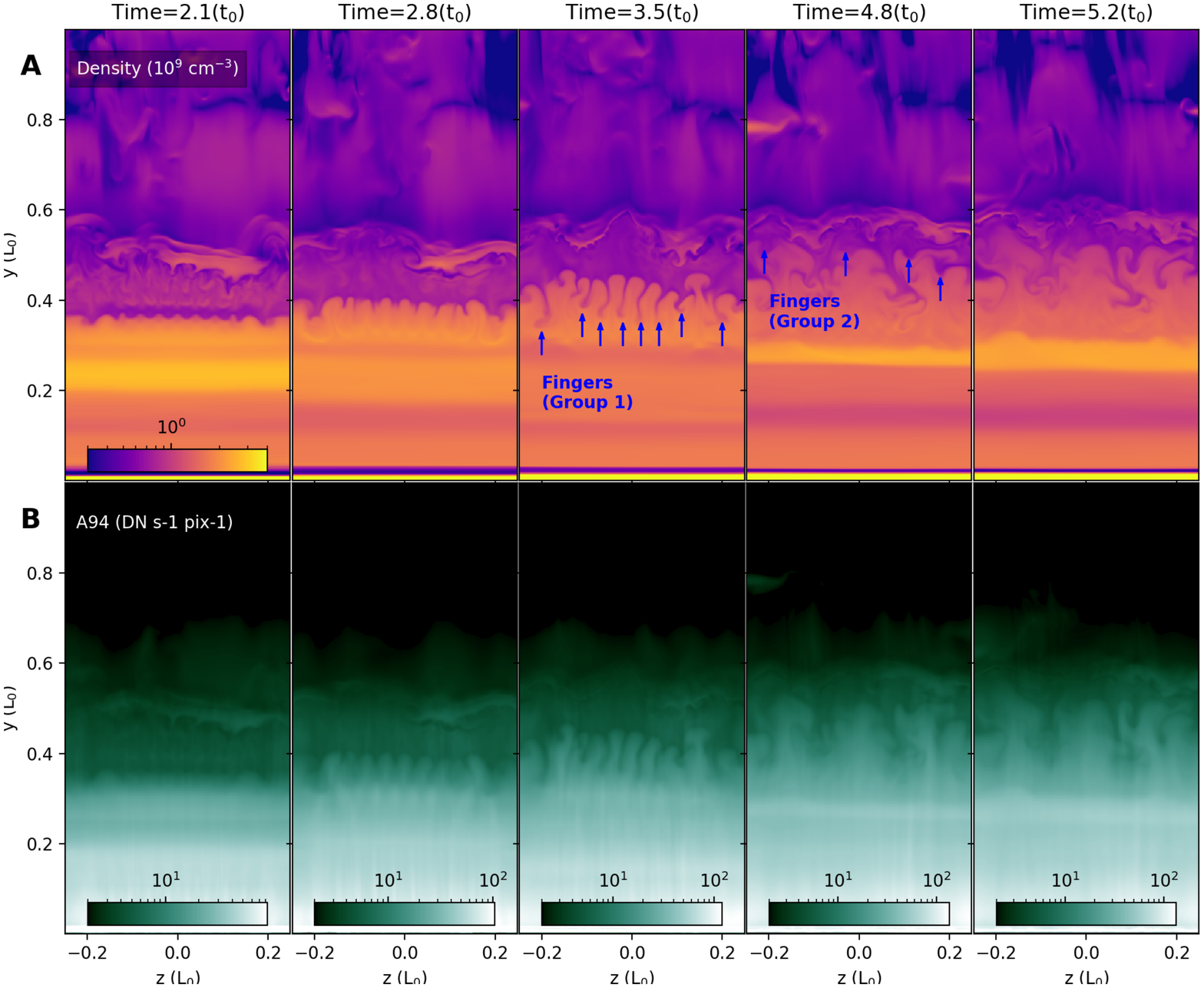}

\noindent {\bf Fig. S3.} \textbf{Temporal evolution of the plasma density and synthetic SDO/AIA 94 \AA\ EUV intensity maps in Case A.} This EUV filter-band is sensitive to $\sim$7 MK flare plasma (through the \ion{Fe}{xviii} line). Arrows indicate Finger-like SAD structures. The finger structures are well-developed until Time=3.5(t$_0$) as marked by ``Group 1", while ``Group 2" shows newly formed fingers at the growing bubble surfaces. 

\newpage
\includegraphics[width=0.8\textwidth]{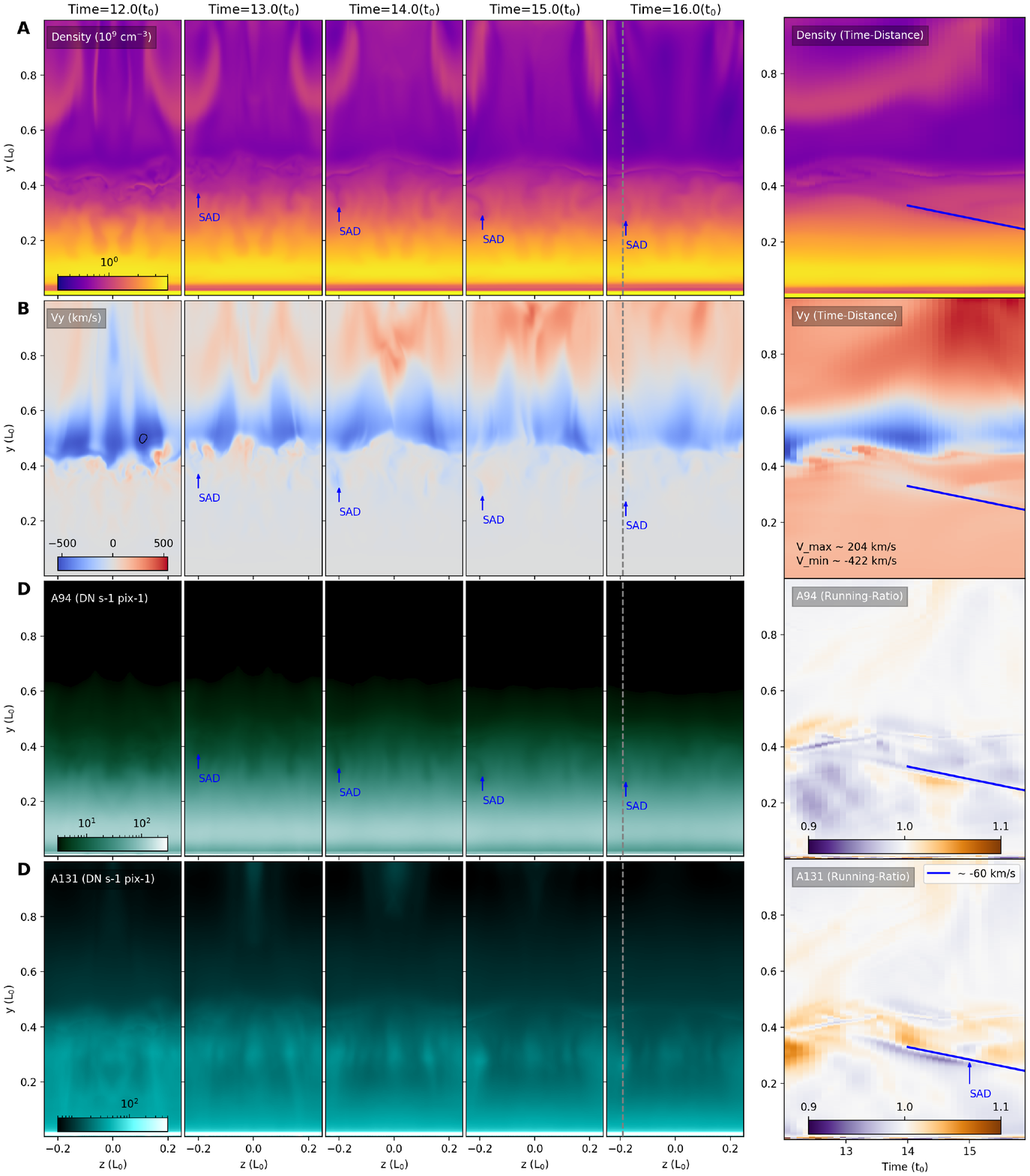}

\noindent {\bf Fig. S4.} \textbf{Temporal evolution of the plasma density, velocity, and synthetic SDO/AIA 94 \AA\ and 131 \AA\ EUV intensity maps in the weak magnetic reconnection regime in Case B.} The AIA running ratio images are obtained along the vertical sampling line ($z$=-0.19\ L$_0$). An example SAD with speed $\sim$ 60 km/s is obtained following the same method as in Fig. 3. 

\newpage
\includegraphics[width=0.80\textwidth]{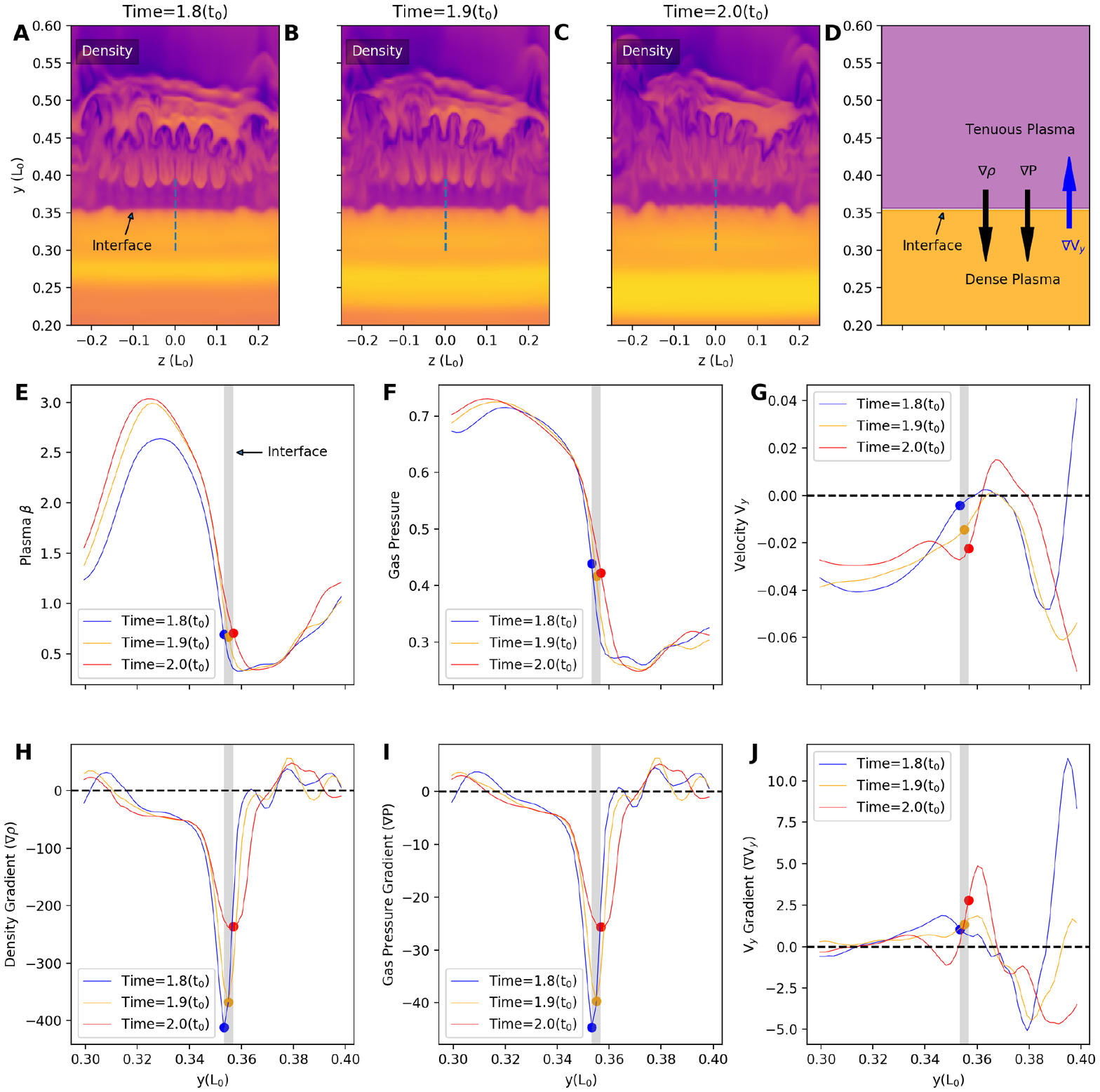}

\noindent {\bf Fig. S5.}
\textbf{Initial development of instabilities at the density interface region}. (A)--(C) Temporal evolution of the plasma density on the center plane ($x=0$). (E)--(G)
Plasma $\beta$, gas pressure, and velocity variation across the interface along the vertical dotted sampling line in (A)--(C). (H)--(J) Density gradient, gas pressure gradient, and velocity (V$_y$ component) gradient along the same sampling line. 
The blue, orange, and red cycles indicate the density interface height at different times 1.8, 1.9, and 2.0$t_0$, respectively.
(D) Schematic representation of the direction of the density gradient, gas pressure gradient, and vertical velocity (V$_y$) gradient.
The gray shadows indicate the location of the density interfaces. The positive V$_y$ gradient near the density interface indicates an upward flow towards the tenuous side from the dense plasma side. Both the density gradient and pressure gradient are downwards, which matches the condition for driving the RTI/RMI well. 

\newpage
\includegraphics[width=0.8\textwidth]{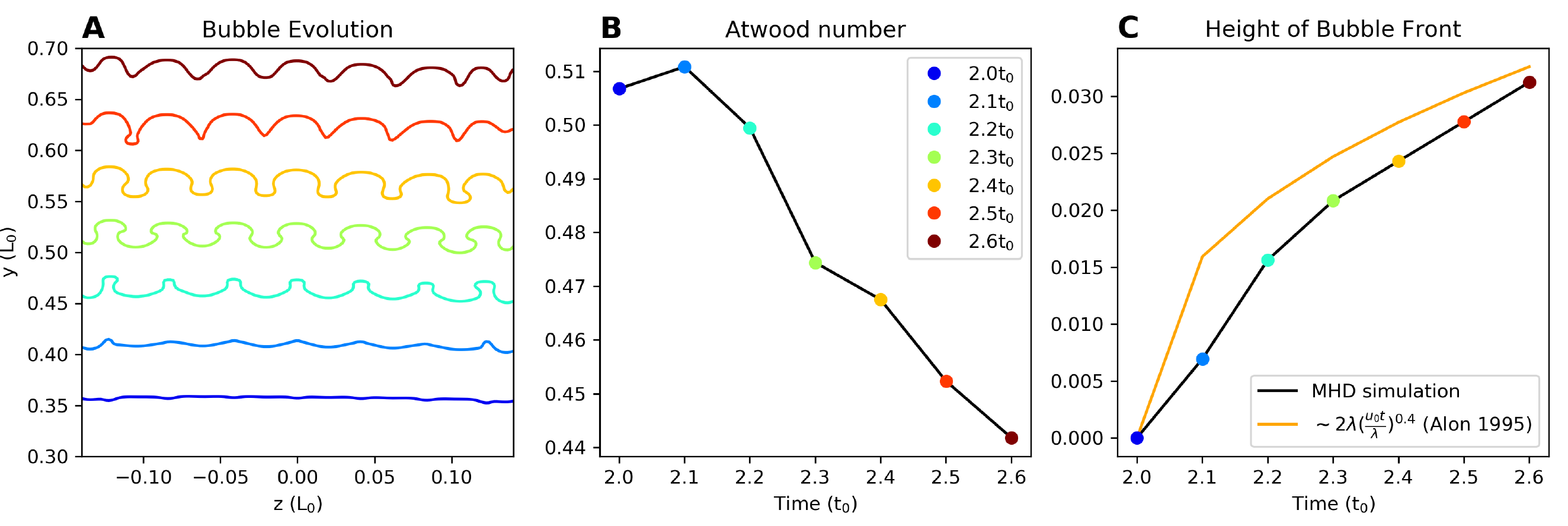}

\noindent {\bf Fig. S6.} \textbf{Bubble evolution during the non-linear RTI/RMI phase.} (A) The bubble interface (contours of the density) during the period 2.0 -- 2.6$t_0$, vertically offset by a fixed interval 0.05L$_0$ for clarity; (B) The average Atwood number $A$ at different times. Here $A = (\rho_{d} - \rho_{t})/(\rho_{d} + \rho_{t})$, where $d$ and $t$ refer to the dense and tenuous plasma, respectively; (C) Time evolution of the height of the bubbles. The black line is from the MHD simulation, and the orange line is based on a theoretical model\cite{Alon1995PhRvL..74..534A}. Here, the scale-invariant power-law index $\theta$ is chosen to be $\sim$ 0.4, the initial perturbation wavelength ($\lambda \sim 0.02$) can be estimated from the panel (A), and the initial perturbation velocity ($u_0$) is $\sim$ 0.02 based on Fig. S5.

\newpage
\includegraphics[width=0.80\textwidth]{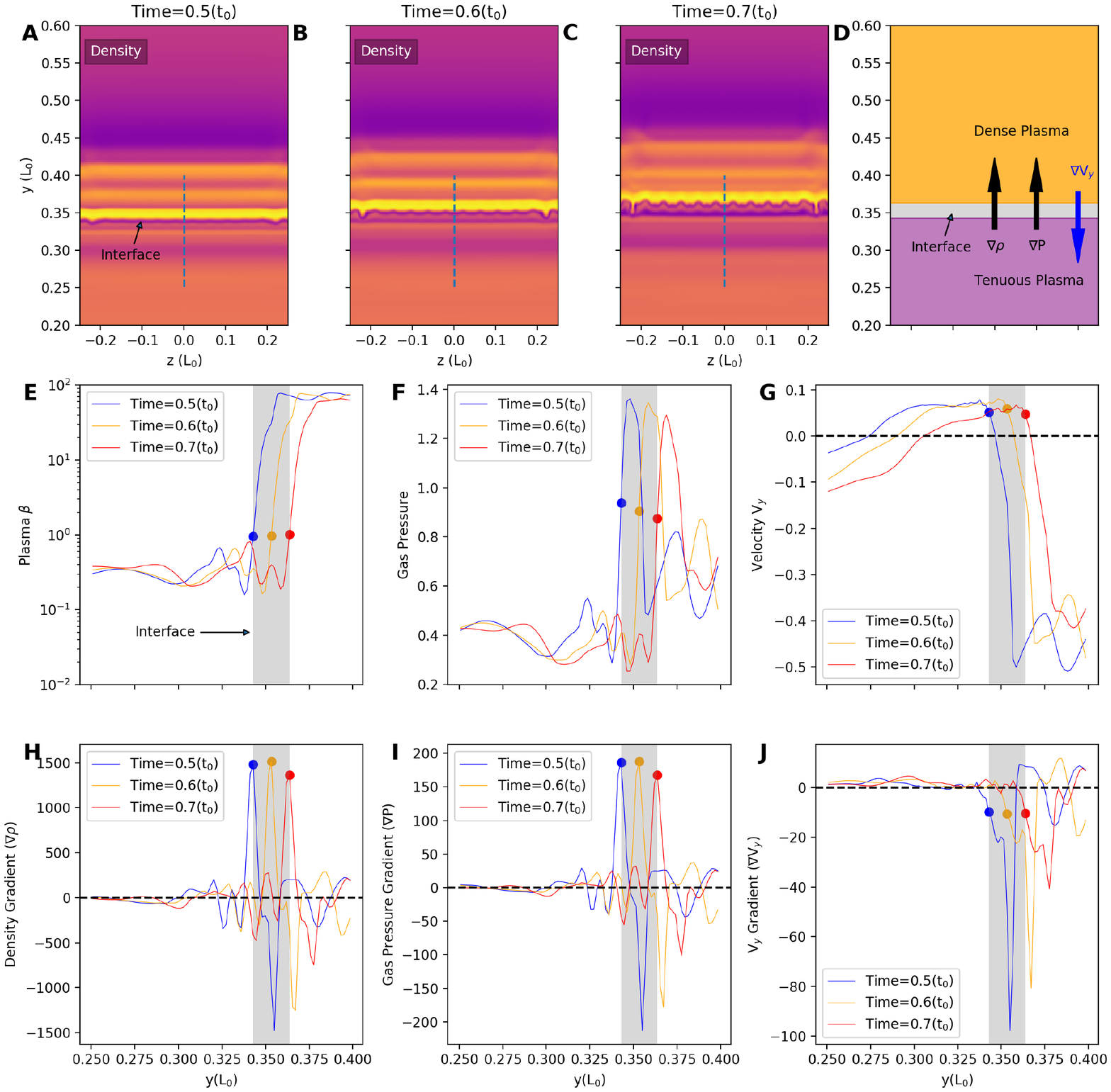}

\noindent {\bf Fig. S7.} \textbf{Development of instabilities at the density interface region with a reversed gradient.} Same as Fig. S5 except that the interface appears at early times with a reversed density structure.

\clearpage





\end{document}